\documentclass{article}

\usepackage[english]{babel}

\usepackage[a4paper,top=2cm,bottom=2cm,left=3cm,right=3cm,marginparwidth=1.75cm]{geometry}

\usepackage{amsmath}
\usepackage{graphicx}
\usepackage[colorlinks=true, allcolors=blue]{hyperref}
\usepackage[auth-sc, affil-it]{authblk}
\usepackage{amssymb}
\usepackage{acro}
\usepackage{comment}
\usepackage{caption}
\usepackage{subcaption}
\usepackage{parskip}
\DeclareAcronym{EEF}{
  short=EEF,
  long=eigenfunction expansion form,
}

\DeclareAcronym{SIFs}{
  short=SIFs,
  long=stress intensity factors,
}

\DeclareAcronym{HOSTs}{
  short=HOSTs,
  long=higher order singular terms,
}

\DeclareAcronym{HORTs}{
  short=HORTs,
  long=higher order regular terms,
}

\DeclareAcronym{SSY}{
  short=SSY,
  long=small scale yielding,
}

\DeclareAcronym{LSY}{
  short=LSY,
  long=large scale yielding,
}

\DeclareAcronym{DIC}{
  short=DIC,
  long=digital image correlation,
}

\DeclareAcronym{FEA}{
  short=FEA,
  long=finite element analysis,
}

\DeclareAcronym{ODM}{
  short=ODM,
  long=over-deterministic method,
}

\DeclareAcronym{FIMTS}{
  short=FIMTS,
  long=modified maximum tangential stress,
}

\DeclareAcronym{MTS}{
  short=MTS,
  long=maximum tangential stress,
}
\DeclareAcronym{MT}{
  short=MT,
  long=middle tension,
}

\DeclareAcronym{SECT}{
  short=SECT,
  long=single-edge cracked tension,
}

\DeclareAcronym{CT}{
  short=CT,
  long=compact tension,
}

\providecommand{\keywords}[1]
{
  \small	
  \textbf{\textit{Keywords---}} #1
}

\title{Advanced crack tip field characterization using conjugate work integrals}
\date{December 13, 2022}
\author[1,*]{David Melching}
\author[1]{Eric Breitbarth}
\affil[1]{German Aerospace Center (DLR), Institute of Materials Research, Linder Hoehe, 51147 Cologne, Germany.}
\affil[*]{Corresponding author: David.Melching@dlr.de}

\usepackage{fancyhdr}
\fancypagestyle{licensefooter}{%
  \fancyhf{}
  
  \fancyfoot[L]{{\footnotesize © 2023. Licensed under CC-BY-NC-ND 4.0 license \url{https://creativecommons.org/licenses/by-nc-nd/4.0/}}}
}

\begin{document}

\maketitle

\thispagestyle{licensefooter}

\begin{abstract}
The quantitative characterisation of crack tip loads is fundamental in fracture mechanics. Although the potential influence of higher order terms on crack growth and stability is known, classical studies solely rely on first order stress intensity factors. We calculate higher order Williams coefficients using an integral technique based on conjugate work integrals and study the convergence with increasing crack tip distance. We compare the integral method to the state-of-the-art fitting method and provide results for higher-order terms with several crack lengths, external forces, and sizes for widely used middle tension, single-edge cracked tension, and compact tension specimen under mode-I loading.
\end{abstract}

\keywords{fracture mechanics, Williams coefficients, higher-order terms, Bueckner integral}

\section{Introduction} \label{sec:intro}

Linear elastic fracture mechanics relies mainly on the \ac{SIFs} related to the first term of \textsc{Williams}' series and represented by the $r^{-1/2}$ stress singularity \cite{Williams1961}. As an infinite series, the \textsc{Williams} \ac{EEF} can have both positive and negative terms \cite{Roux-Langlois2015}. For pure mode I (cf. Section \ref{sec:method_williams}), the \ac{EEF} for the displacement field reads: 
\begin{align}
u_{i}(r,\theta) &= \underbrace{\sum_{n=-\infty}^{-1} A_{n} g_{{\rm I},i}(\theta, n) \ r^{\frac{n}{2}}}_{\text{supersingular terms}} + \underbrace{A_0 g_{{\rm I},i}(\theta, 0)}_{\text{translations}} + \underbrace{A_1 g_{{\rm I},i}(\theta, 1) \sqrt{r}}_{\text{singular}} + \underbrace{\sum_{n=2}^{\infty} A_n g_{{\rm I},i}(\theta, n) \ r^{\frac{n}{2}}}_{\text{subsingular terms}} \label{eq:williams_displ_full}
\end{align}
The influence of higher order terms ($n\le-1$ or $n\ge2$) on the stress field of a cracked plate under in-plane loading can be important in some engineering applications \cite{Berto2010,HadjMeliani2011,Berto2011,Berto2013,Stepanova2022}. The further \ac{FIMTS} criterion takes into account several stress terms and can assess the mode I fracture toughness ($P_{\mathrm{Ic}}$) of rocks better than the traditional \ac{MTS} criterion \cite{Wei2018}. Furthermore, the specimen geometry not only influences the \ac{SIFs} but also the stress biaxiality \cite{Leevers1982} and can have a significant effect on fatigue crack growth rates \cite{Varfolomeev2011,Seitl2008}. Studies indicate that in many cases at least seven terms are needed to reconstruct the stress field at the crack tip with a sufficiently small error \cite{Vesely2016}.

Negative terms with $n\le-1$ are known as \ac{HOSTs}, outside expansions or supersingular terms \cite{Chen2003, Chen1997, Roux-Langlois2015}. The role of these terms is contradictory. On the one hand, they should be omitted to ensure that the strain energy as well as the displacements stay bounded in the near-tip region \cite{Chen1997}. On the other hand, the incorporation of \ac{HOSTs} seems to improve the description of the non-linearity of the process zone for \ac{SSY} \cite{Hui1995,Chen1997}. In case of \ac{LSY}, \ac{HOSTs} effect the $J$-Integral due to the interaction between singular and regular higher terms \cite{Jeon2001}.

Positive terms with $n\ge2$ are called \ac{HORTs}, subsingular terms, or higher order non-singular terms. They are useful to match the near field with remote geometry or boundary conditions \cite{Baldi2022}. The first regular stress term ($n=2$), called $T$-stress, acts parallel to the crack and was investigated in more detail within multi-parameter approaches  \cite{Shlyannikov2014,Baptista2020,Tong2002,Shaikeea2022,Berto2013}. The $T$-stress is related to crack path stability, plastic zone shape, and constraint parameters of the plastic zone size. Its sign determines shielding or anti-shielding behaviour of the plastic zone shape and size at the crack tip \cite{Gupta2015}. Positive $T$-stress reduces, while negative $T$-stress increases the size of the plastic zone \cite{Sobotka2011}. 
Consequently, a positive $T$-stress leads to directional instability associated with the possibility of crack path deflection. Derived from this, there are models for crack deflection under biaxial loading that are based on the hypothesis of maximum circumferential stress and, in addition to $K_{\mathrm{I}}$ and $K_{\mathrm{II}}$, also take $T$-stress into account \cite{Chen2002,Matvienko2012}.
Furthermore, the third regular term ($n=3$) is responsible for the stability of the crack propagation and the fourth term characterises the maximum shear stress as a function of the distance from the crack tip in the crack propagation direction \cite{Cotterell1966}. 

Higher order terms are particularly important for matching the experimental results obtained with \ac{DIC} with the theoretical near-field to determine \ac{SIFs}. Truncated \textsc{Williams} series can be projected directly onto the measured field to determine the crack tip position and \ac{SIFs}. To increase accuracy, singular and regular terms (e.g. $-3 \le n \le 7$) are considered \cite{Yates2010,Roux-Langlois2015,Rethore2015}.  In the presence of plasticity, the shape of the plastic zone is mainly given by the field $n=-3$, but can also be approximated by $K_{\mathrm{I}}$ and $T$ if all fields except $n=-3$ are considered \cite{Yates2010}. 

Only under certain assumptions and simplifications, the \textsc{Williams} coefficients can be determined analytically \cite{Hello2012}. There are also several special finite element approaches for calculating the \textsc{Williams} coefficients in numerical simulations \cite{Xiao2007,Chidgzey2005,Treifi2009,Tsang2008}. \textsc{Ayatollahi} \& \textsc{Nejati} \cite{Ayatollahi2011} introduced the so-called \ac{ODM} to determine higher-order terms based on fitting the \ac{EEF} to given displacement data. Later, Malikova \& Vesely \cite{Malikova2015} used this approach to study the influence of higher order terms on the stress field. However, the \ac{ODM} has two main limitations. First, the optimization can fail to converge if the initial values are too far from the actual optimum. Secondly, and most crucially its results depend on the choice how many terms of the series are taken into account for the ansatz function (see Section \ref{sec:method_odm} below). A change in the number of considered terms alters the results for all coefficients.

Well-known integral techniques like the $J$- or interaction integrals are reliable methods for calculating $K_{\rm I}$ and $K_{\rm II}$ based on given displacement and stress fields from \ac{FEA} or \ac{DIC} \cite{Stern1976,Rice1968,Molteno2015}.
The pseudo-orthogonal property of the \ac{EEF} allows to establish invariant integrals and weight functions for many crack problems. Based on Betti’s reciprocal theorem, \textsc{Bueckner} \cite{Bueckner1973} proposed a work conjugate integral. \textsc{Chen} \cite{Chen1985,CH94} showed that \textsc{Bueckner}'s integral is path-independent and can be used to determine all coefficients of the \textsc{Williams} series, including the \ac{SIFs} and $T$-stress. Actually, the $J$- and $M$-integral are special cases of \textsc{Bueckner}’s integral \cite{Chen2003}. However, there exists no method to calculate higher-order terms from $J$ or $M$.

While there exist several methods to calculate \ac{SIFs} and corresponding studies are widely available for common specimen geometries, they are rarely present for higher order terms \cite{Hello2012,Chen1997}. In particular, to the best of our knowledge, there exists no scientific study of higher-order terms using the mentioned integral method.

In this work, we address this gap by calculating higher order terms for several standard specimen geometries and crack lengths. In contrast to previous studies, we use \textsc{Chen}'s integral method based on \textsc{Bueckner}'s conjugate work integral and compare its performance to the classical \ac{ODM} (Section \ref{sec:results_method_comp}). We focus on \ac{HORTs} as there is no \ac{SSY} in our linear-elastic finite element simulations \cite{Chen1997}. Furthermore, we rigorously investigate the relative distance from the crack tip which is needed for convergence of \ac{HORTs} depending on element and integral mesh size (Section \ref{sec:results_conv_study}). Finally, we carry out extensive parameter studies by calculating \ac{HORTs} up to order $n=5$ for \ac{MT}, \ac{SECT} and \ac{CT} specimen under mode-I loading conditions for various specimen dimensions, crack lenghts, and loading forces. 

The results provide an effective way for compressing crack tip fields in single parameters to generate a rich database for further data-driven analysis. This enables a structured investigation of effects between specimen geometry, loading conditions, crack tip field, crack propagation and crack path stability.
Both the \textsc{Bueckner-Chen} integral method and the over-deterministic method were implemented in Python. The implementation is a part of the software project \textit{crackpy} \cite{crackpy-github}. It enables researchers and users to analyze crack propagation by characterizing crack tip fields during experiments or simulations automatically.

\section{Analytical methods} \label{sec:methods_ana}
\begin{figure}
    \centering
    \includegraphics[width=0.8\textwidth]{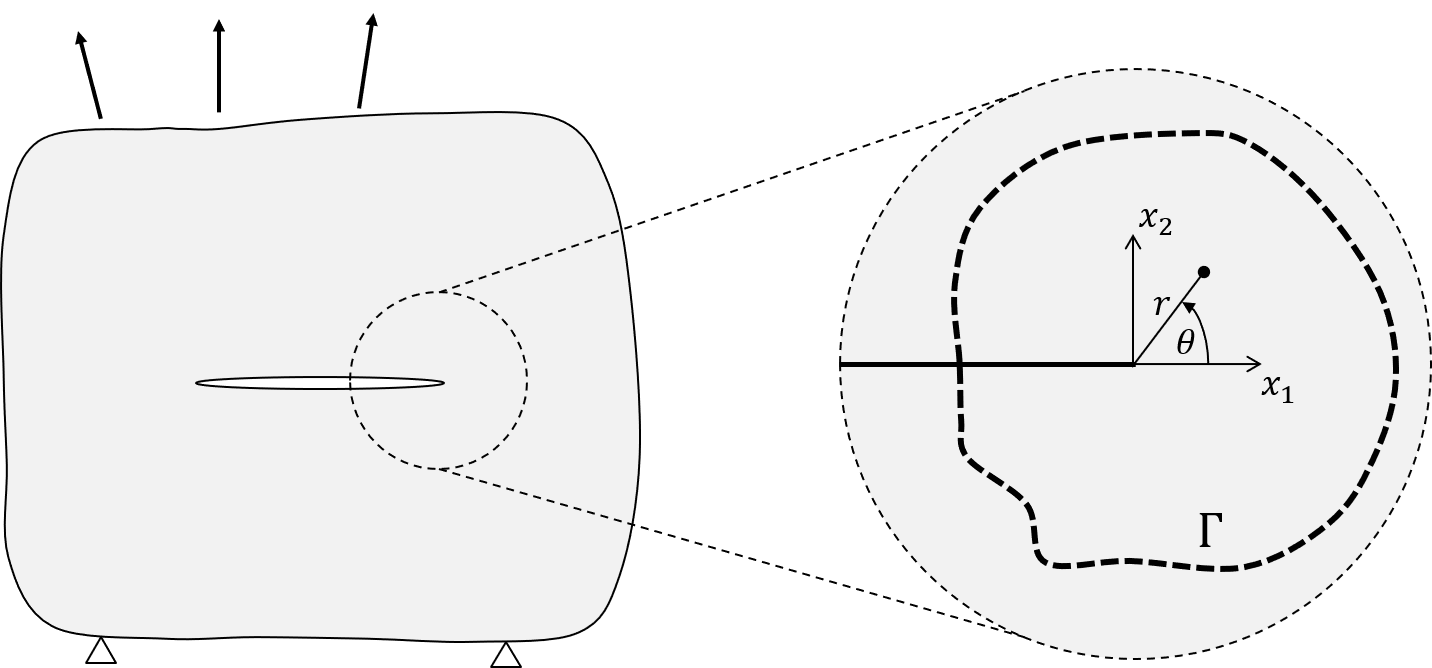}
    \caption{Local cartesian and polar coordinate systems at the crack tip. Integration path $\mathrm{\Gamma}$ surrounding the crack tip.}
    \label{fig:local_coordinates}
\end{figure}

We consider a cracked plate subjected to an arbitrary in-plane load and introduce a local cartesian coordinate system $\left(x_1,x_2\right)$, where the line $x_1\le 0$ is occupied by the crack path. We define local polar coordinates $\left(r,\theta\right)$, where $r\ge 0,\theta\in\left[-\pi,\pi\right]$ and
\begin{align} \label{eq:polar_coors}
\begin{split}
    x_1=r\cos(\theta), \\ 
    x_2=r\sin(\theta),
\end{split}
\end{align}
see Figure \ref{fig:local_coordinates}.

\subsection{Williams series} \label{sec:method_williams}
In planar linear-elasticity the stress and displacement fields can be written in polar coordinates $(r, \theta)$ (see Equation \ref{eq:polar_coors} and Figure \ref{fig:local_coordinates}) around the crack tip as \cite{Williams1961,Kuna}:
\begin{align}
\sigma_{ij}(r,\theta) &= \sum_{n=1}^{\infty}{r^{\frac{n}{2}-1}\ \left( A_n f_{\text{I},ij}(\theta,n) + B_n f_{\text{II},ij}(\theta,n) \right)}, \label{eq:williams_stress} \\
u_{i}(r,\theta) &= \sum_{n=0}^{\infty} \frac{r^{\frac{n}{2}}}{2\mu} \ \left( A_n g_{\text{I},i}(\theta,n) + B_n g_{\text{II},i}(\theta,n) \right). \label{eq:williams_displ}
\end{align}

The parameters $A_n, B_n \in \mathbb{R}$ are called \textsc{Williams} coefficients and depend on the loading conditions. The trigonometric eigenfunctions $f_{\text{I},ij},f_{\text{II},ij}, g_{\text{I},i},g_{\text{II},i}$ for the stress tensor are defined as follows:
\begin{equation}
f_{\text{I},11}(\theta,n)=\frac{n}{2} \left\{ \left[ 2+(-1)^n+\frac{n}{2} \right] \cos \left(\left(\frac{n}{2}-1\right) \theta \right) - \left( \frac{n}{2}-1 \right) \cos \left(\left(\frac{n}{2}-3\right) \theta \right) \right\}
\end{equation}

\begin{equation}
f_{\text{II},11}(\theta,n)=\frac{n}{2} \left\{ \left[ -2+(-1)^n-\frac{n}{2} \right] \sin \left(\left(\frac{n}{2}-1\right) \theta \right) + \left( \frac{n}{2}-1 \right) \sin \left(\left(\frac{n}{2}-3\right) \theta \right) \right\}
\end{equation}

\begin{equation}
f_{\text{I},22}(\theta,n)=\frac{n}{2} \left\{ \left[ 2-(-1)^n-\frac{n}{2} \right] \cos \left(\left(\frac{n}{2}-1\right) \theta \right) + \left( \frac{n}{2}-1 \right) \cos \left(\left(\frac{n}{2}-3\right) \theta \right) \right\}
\end{equation}

\begin{equation}
f_{\text{II},22}(\theta,n)=\frac{n}{2} \left\{ \left[ -2-(-1)^n+\frac{n}{2} \right] \sin \left(\left(\frac{n}{2}-1\right) \theta \right) - \left( \frac{n}{2}-1 \right) \sin \left(\left(\frac{n}{2}-3\right) \theta \right) \right\}
\end{equation}

\begin{equation}
f_{\text{I},12}(\theta,n)=f_{\text{I},21}(\theta,n)=\frac{n}{2} \left\{ \left( \frac{n}{2} - 1 \right) \sin \left(\left(\frac{n}{2}-3\right) \theta \right) - \left[ \frac{n}{2}+(-1)^n \right] \sin \left(\left(\frac{n}{2}-1\right) \theta \right) \right\}
\end{equation}

\begin{equation}
f_{\text{II},12}(\theta,n)=f_{\text{II},21}(\theta,n)=\frac{n}{2} \left\{ \left( \frac{n}{2} - 1 \right) \cos \left(\left(\frac{n}{2}-3\right) \theta \right) - \left[ \frac{n}{2}-(-1)^n \right] \cos \left(\left(\frac{n}{2}-1\right) \theta \right) \right\}
\end{equation}
and the functions for the displacement part are defined as

\begin{equation}
g_{\text{I},1}(\theta,n)=\left[ \kappa + (-1)^n + \frac{n}{2} \right] \cos \left( \frac{n}{2} \theta \right) - \frac{n}{2} \cos \left( \left( \frac{n}{2} - 2 \right) \theta \right)
\end{equation}

\begin{equation}
g_{\text{II},1}(\theta,n)=\left[ -\kappa + (-1)^n - \frac{n}{2} \right] \sin \left( \frac{n}{2} \theta \right) + \frac{n}{2} \sin \left( \left( \frac{n}{2} - 2 \right) \theta \right)
\end{equation}

\begin{equation}
g_{\text{I},2}(\theta,n)=\left[ \kappa - (-1)^n - \frac{n}{2} \right] \sin \left( \frac{n}{2} \theta \right) + \frac{n}{2} \sin \left( \left( \frac{n}{2} - 2 \right) \theta \right)
\end{equation}

\begin{equation}
g_{\text{II},2}(\theta,n)=\left[ \kappa + (-1)^n - \frac{n}{2} \right] \cos \left( \frac{n}{2} \theta \right) + \frac{n}{2} \cos \left( \left( \frac{n}{2} - 2 \right) \theta \right)
\end{equation}

Here, $\kappa$ is the \textsc{Kolosov} constant, namely, $\kappa = (3-\nu) / (1+\nu)$ under plane stress conditions and $\kappa=3-4\nu$ for plane strain, where $\nu>0$ is the \textsc{Poisson} ratio and $\mu = E/(2(1+\nu))$ is the shear modulus.

\subsection{Stress intensity factors and rigid body motion} \label{sec:method_sifs}
The stress intensity factors $K_{\rm I}$ and $K_{\rm II}$ are proportional to the first order coefficients $A_1$ and $B_1$, respectively. The $T$-stress acting upon the crack’s longitudinal axis is proportional to the second order coefficient $A_2$. More precisely,
\begin{align}
K_{\rm I} &= \sqrt{2\pi} A_1, \\
K_{\rm II} &= -\sqrt{2\pi} B_1, \\
T &= 4 A_2.
\end{align}
We recall that the term $n=0$ in the displacement expansion
\begin{align}
    \begin{pmatrix}
        u_1 \\ 
        u_2
    \end{pmatrix}
    = \frac{\kappa + 1}{2\mu} 
    \begin{pmatrix}
        A_0 \\
        B_0
    \end{pmatrix}
\end{align}
corresponds to a translation. Similarly, the second mode-II expansion term corresponds to a rotation with angle  $\alpha = -(\kappa +1)B_2/(2\mu)$ around the crack tip. The rotation component is therefore expressed as follows,
\begin{align}
    \begin{pmatrix}
        u_1 \\ 
        u_2
    \end{pmatrix}
    = \frac{\kappa + 1}{2\mu} B_2 
    \begin{pmatrix}
        -r\sin\theta  \\
        r\cos\theta
    \end{pmatrix}.
\end{align}
Thus, the terms $A_0, B_0$ and $B_2$ define a stress-free rigid body motion.

\subsection{Determination of \ac{HORTs}} \label{sec:method_det_will_coeffs}
Our goal is to determine the higher order regular terms (HORTs) $(A_n, B_n)_{n \ge 1}$ for a given load case resulting in observed (measured) displacements and stresses $u_i, \sigma_{ij}$. 

In this paper, we follow an invariant integral approach for the determination of the higher order terms based on \textsc{Bueckner}'s conjugate work integral \cite{Bueckner1973}. Let $\sigma_{ij}^{(1)}, \sigma_{ij}^{(2)}$ and $u_{i}^{(1)}, u_{i}^{(2)}$ be stresses and displacement fields corresponding to two load conditions for the same plane, linear-elastic crack configuration. We consider the \textsc{Bueckner} integral
\begin{equation} \label{eq:Bueckner_integral}
    I_{\Gamma} = \int_\Gamma \left( \sigma_{ij}^{(1)} u_{i}^{(2)} - \sigma_{ij}^{(2)} u_{i}^{(1)}\right)n_j \ \text{d}s,
\end{equation}
where $\Gamma$ denotes a closed path around the crack tip (see Figure \ref{fig:local_coordinates}) and $n=(n_1,n_2)$ stands for the (outward) normal vector with respect to $\Gamma$. In case of traction-free crack faces, path-independence of $I_\Gamma$ is a direct consequence of Betti's reciprocity theorem.

Moreover, \textsc{Chen} \cite{Chen1985} proved that all terms of the \textsc{Williams} series can be determined using $I_{\Gamma}$. We follow \textsc{Chen}'s proof to derive a formula for $A_n$ and $B_n$: Let $(\sigma_{ij}^{(1)}, u_{i}^{(1)}) = (\sigma_{ij}, u_i)$ belong to a given load case for which we aim to calculate the \ac{HORTs}. Then, we introduce a second virtual test load case $(\sigma_{ij}^{(2)}, u_{i}^{(2)}) = (\hat{\sigma}_{ij}, \hat{u}_i)$ in Equation \ref{eq:Bueckner_integral} and write down the \ac{EEF} for both as
\begin{align}
\sigma_{ij}(r,\theta) &= \sum_{n=-\infty}^{\infty}{r^{\frac{n}{2}-1}\ \left( A_n f_{\text{I},ij}(\theta,n) + B_n f_{\text{II},ij}(\theta,n) \right)}, \\
u_{i}(r,\theta) &= \sum_{n=-\infty}^{\infty} \frac{r^{\frac{n}{2}}}{2\mu} \ \left( A_n g_{\text{I},i}(\theta,n) + B_n g_{\text{II},i}(\theta,n) \right), \\
\hat\sigma_{ij}(r,\theta) &= \sum_{m=-\infty}^{\infty}{r^{\frac{m}{2}-1}\ \left( \hat A_m f_{\text{I},ij}(\theta,m) + \hat B_m f_{\text{II},ij}(\theta,m) \right)}, \\
\hat u_{i}(r,\theta) &= \sum_{m=-\infty}^{\infty} \frac{r^{\frac{m}{2}}}{2\mu} \ \left( \hat A_m g_{\text{I},i}(\theta,m) + \hat B_m g_{\text{II},i}(\theta,m) \right).
\end{align}

Inserting the expansions into Equation \ref{eq:Bueckner_integral}, \textsc{Bueckner}'s integral reads
\begin{equation}
    I_{\Gamma} = \sum_{n,m} I_\Gamma(n,m),
\end{equation}
where due to orthogonality property of the eigenfunctions of the crack problem \cite{Chen1985}
\begin{equation}
    I_{\Gamma}(n,m)=\begin{cases}
    -\frac{\pi(\kappa+1)}{\mu}{(-1)}^{n+1}n(A_n{\hat{A}}_m+B_n{\hat{B}}_m), & \text{if }m+n=0, \\
    0, & \text{else.}
    \end{cases}
\end{equation}

In order to determine the regular term $A_n$ for arbitrary $n\ge 1$, we choose the test load case belonging to the corresponding singular term. More precisely, to compute $A_n$ we choose ${\hat{A}}_m=\delta_{m,-n},\ {\hat{B}}_m=0$ and to derive $B_n$, we choose the test ${\hat{B}}_m=\delta_{m,-n},\ {\hat{A}}_m=0$ (Kronecker delta notation). These two choices lead to integrals $I_\Gamma^\mathrm{I}\left(n\right)$ and $I_\Gamma^{\mathrm{II}}\left(n\right)$, respectively. A simple rearrangement yields
\begin{align}
    A_n &= - \frac{\mu}{\kappa + 1} \frac{1}{\pi n(-1)^{n+1}} I_\Gamma^{\rm I}(n), \\
    B_n &= - \frac{\mu}{\kappa + 1} \frac{1}{\pi n(-1)^{n+1}} I_\Gamma^{\rm II}(n).
\end{align}
Throughout this work, we use the physical units MPa for stress and mm for length. A simple consideration reveals that $A_n$ and $B_n$ have the unit $\text{MPa} \cdot \text{mm}^{1-n/2}$.

\section{Numerical methods} \label{sec:method_simulations}
We demonstrate and analyze the proposed integral method on three specimen geometries under pure mode-I loadings:

\begin{enumerate}
    \item center-cracked under middle tension (MT)
    \item single-edge crack under uniaxial tension (SECT)
    \item compact tension specimen (CT)
\end{enumerate}

\begin{figure}
     \centering
     \begin{subfigure}[b]{0.3\textwidth}
         \centering
         \includegraphics[width=\textwidth]{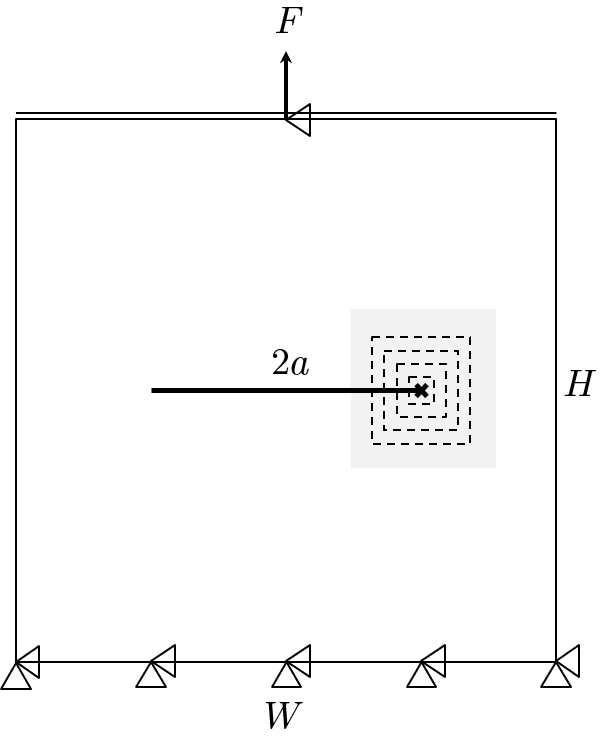}
         \caption{MT}
         \label{fig:MT_specimen}
     \end{subfigure}
     \hfill
     \begin{subfigure}[b]{0.2\textwidth}
         \centering
         \includegraphics[width=0.85\textwidth]{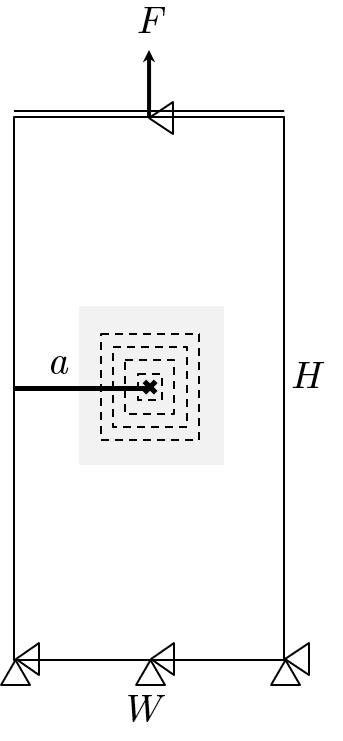}
         \caption{SECT}
         \label{fig:SECT_specimen}
     \end{subfigure}
     \hfill
     \begin{subfigure}[b]{0.35\textwidth}
         \centering
         \includegraphics[width=\textwidth]{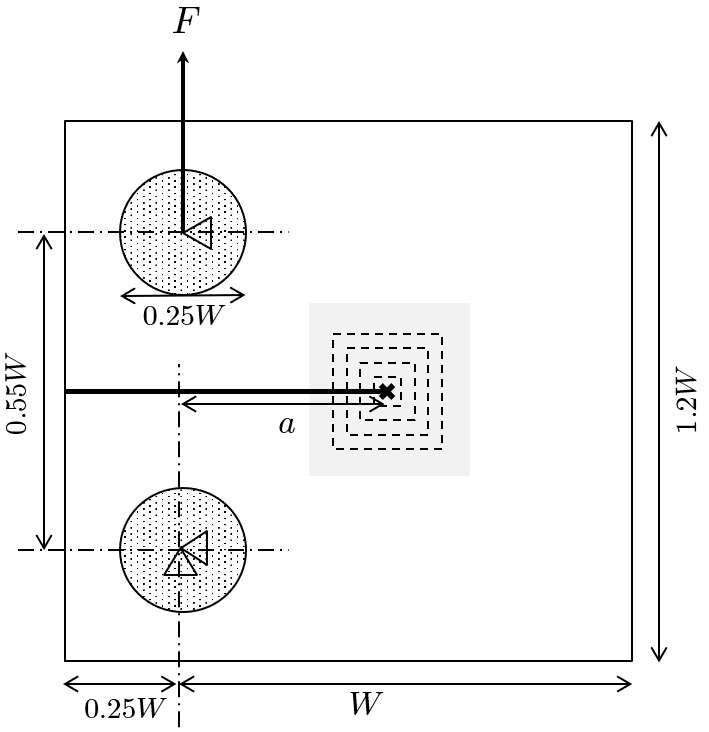}
         \caption{CT}
         \label{fig:CT_specimen}
     \end{subfigure}
    \caption{Specimen geometries with crack length $a$ [mm], width $W$ [mm], height $H$ [mm], and force $F$ [N]. Dashed lines indicate the integral evaluation paths and the areas where the mesh size $l_{\mathrm{e}}$ was prescribed are shown as gray boxes around the crack tip. The plate thickness in all cases is $t=2$ mm.}
    \label{fig:specimen_geometries}
\end{figure}

The specimens were modelled and simulated as a 2D model with PLANE182 elements using the software Ansys\textsuperscript{\textregistered} Mechanical \cite{Ansys}. Here, an unstructured mesh with element size 2 mm was used. The mesh was refined to a mesh size $l_{\mathrm{e}}$ [mm] inside the integration area around the crack tip. We used a plate thickness of $t=2$ mm and the plane stress condition (KEYOPT,1,3,3) for all simulations. In order to closely match the material properties of the standard alloy AA2024-T3, we took a Poisson ratio of $\nu= 0.33$ and a Young modulus of $E=72000$ MPa for all simulations. Figure \ref{fig:specimen_geometries} shows the exact specimen geometries and boundary conditions. 

In order to emulate the situation of \ac{DIC} measurements, we export and save the nodal coordinates, displacements, and strains rather than calculating the \textsc{Bueckner} integral or fitting the \textsc{Williams} coefficients directly on finite elements. The nodal stresses are calculated using Hooke’s linear elasticity law.

\subsection{\textsc{Bueckner-Chen} integral method} \label{sec:method_int}
\begin{figure}
    \centering
    \includegraphics[width=\textwidth]{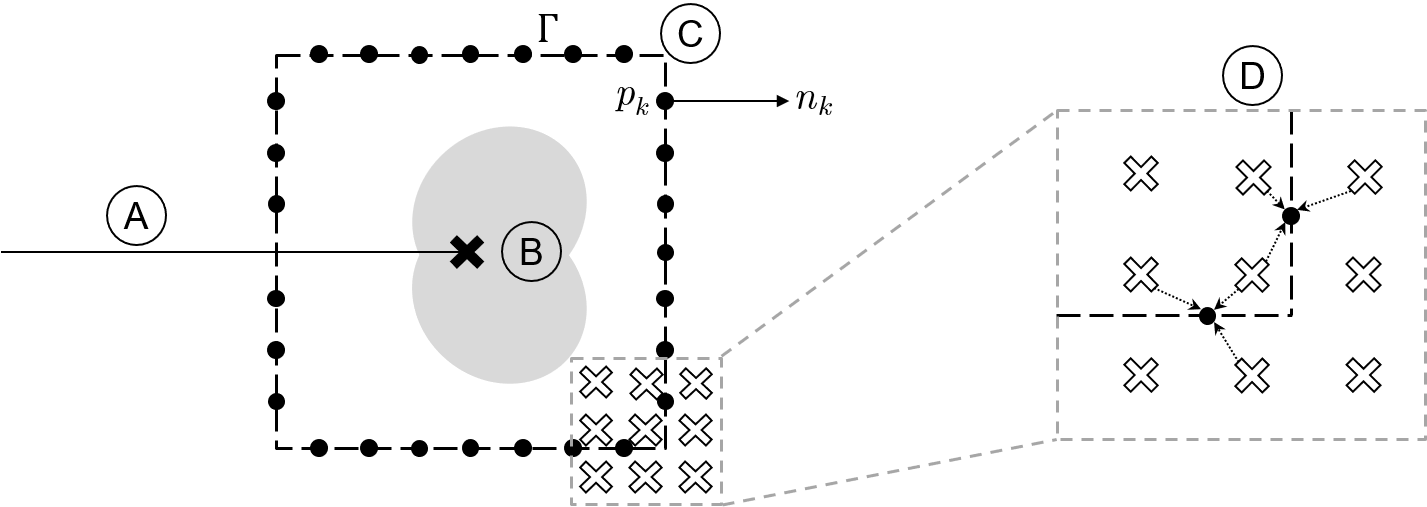}
    \caption{\textsc{Bueckner-Chen} integral method (INT): Crack path (A), crack tip and crack tip field (B), and integration path $\Gamma$ approximated by integration points $p_k$ with corresponding normal vector $n_k$ (C). The nodal displacement data (indicated by white X’s) is interpolated onto the integral grid (D).}
    \label{fig:int_method}
\end{figure}

For the numerical approximation of $I_\Gamma$, we discretize 
\begin{equation*}
    \Gamma \approx \Gamma_P=\{p_1,p_2,\dots,p_P\},
\end{equation*}
where $P$ denotes the number of equidistant integration points (see Figure \ref{fig:int_method}). Note that instead of using the nodal information directly, we first interpolate it on this structured grid. Therefore, the method is independent of the finite element mesh and can easily be applied to real experimental data as well. For definiteness and simplicity, we choose the integration paths to be square-shaped and centered at the crack tip. In particular, this choice allows for a very simple implementation of the standard normal $n$. The nodal displacements, strains, and stresses are interpolated (piecewise linear) onto the integral grid points from which the discretized integral is calculated as described in Section 2.3 to determine the coefficients $A_n,B_n$.

\subsection{Over-deterministic method} \label{sec:method_odm}
The idea of the over-deterministic method, is to optimize the coefficients of the \textsc{Williams} series expansion of order $N$, i.e. the truncated Williams series
\begin{align}
\sigma_{ij}^N(r,\theta) &= \sum_{n=1}^{N}{r^{\frac{n}{2}-1}\ \left( A_n f_{\text{I},ij}(\theta,n) + B_n f_{\text{II},ij}(\theta,n) \right)}, \label{eq:williams_stress_expansion} \\
u_{i}^N(r,\theta) &= \sum_{n=0}^{N} \frac{r^{\frac{n}{2}}}{2\mu} \ \left( A_n g_{\text{I},i}(\theta,n) + B_n g_{\text{II},i}(\theta,n) \right). \label{eq:williams_displ_expansion}
\end{align}
in order to match the finite element simulation $u_i^{\mathrm{FE}}$ on a large number of points $p_k$. Therefore, it boils down to the minimization task:
\begin{equation} \label{eq:odm_minimization}
    \left(A_1,\dots,A_N,B_1,\dots,B_N\right) \mapsto \sum_{k=1}^{P}\sum_{i=1}^{2}\left(u_i^N\left(p_k\right)-u_i^{\mathrm{FE}} \left(p_k\right)\right)^2 \longrightarrow \mathrm{min}
\end{equation}

\begin{figure}
    \centering
    \includegraphics[width=\textwidth]{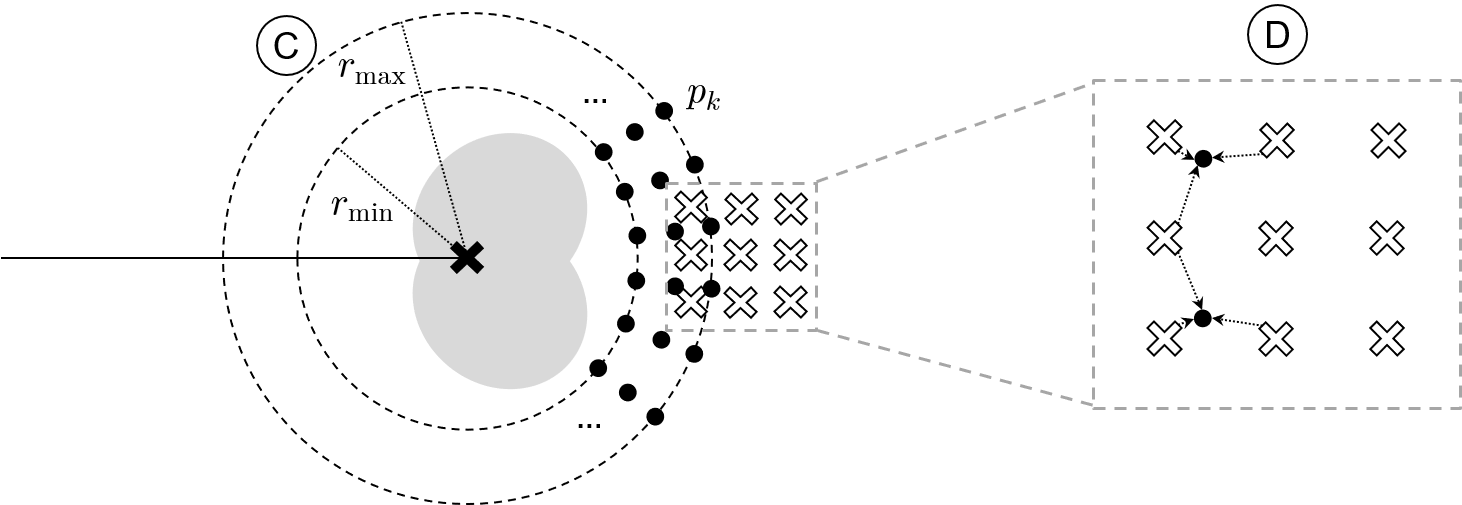}
    \caption{Over-deterministic method (ODM): We choose an annular fitting domain (C) with inner radius $r_{\mathrm{min}}$ and outer radius $r_{\mathrm{max}}$. The fitting domain is equidistantly discretized. The nodal finite element data (indicated by white X’s) is interpolated onto the fitting domain grid (D).}
    \label{fig:odm}
\end{figure}

We remark that a change of the order $N$ alters all calculated coefficients $A_n, B_n$. In order to guarantee a faithful comparison of methods, in contrast to the method introduced in \cite{Ayatollahi2011}, we interpolate the exported displacement data onto a grid in the fitting domain rather than working directly with the finite element nodes. The domain of optimization, called \textit{fitting domain}, is given by an annulus of inner and outer radii $r_{\mathrm{min}}$ and $r_{\mathrm{max}}$, respectively. In polar coordinates $(r,\theta) \in [r_{\mathrm{min}},r_{\mathrm{max}}] \times [-\pi,\pi]$ the fitting domain is discretized by a regular grid of tick size $\delta > 0$ leading to $P = P(\delta) = 2\pi\left(r_{\mathrm{max}}-r_{\mathrm{min}}\right) \delta^{-2}$ fitting points (see Figure \ref{fig:odm}).

\section{Results and discussion} \label{sec:results}
In Section \ref{sec:results_conv_study}, we study the convergence of the \textsc{Bueckner-Chen} integral method in terms of the mesh size $l_{\mathrm{e}}$, the number of integration points $P$ and the integration distance from the crack tip measured in distance radii $r$ relative to the crack length $a$. Then, in Section \ref{sec:results_method_comp}, we compare the approximation error of the proposed integral method with the over-deterministic method. Finally, in Section \ref{sec:results_param_study}, we carry out a full parameter study on the crack length $a$, the external traction force $F$, and the width and height $W, H$ of the specimen. 

\subsection{Convergence study} \label{sec:results_conv_study}
To confirm the path-independence of the $I_\Gamma$-integral we investigate the numerical approximation method introduced above. It is expected that the closer the distance of the integration path to the crack tip the more inaccurate the results are due to the singular behavior at the crack tip field \cite{Ayatollahi2011}. Although the \ac{HORTs} with $n>2$ vanish for $r\rightarrow 0$, the corresponding test functions are divergent with a singularity scaling of $r^{-n/2}$ and $r^{-n/2-1}$ for the test displacements and strains, respectively. This leads to numerical issues close to the crack tip when calculating higher order terms.

To investigate the convergence behavior, we choose a squared MT-specimen (see Figure \ref{fig:specimen_geometries}) with height $H=W=200\ \mathrm{mm}$, a crack length of $a=50\ \mathrm{mm}$ ($\alpha=0.5$), and a force of $F=10\ \mathrm{kN}$. We compute the coefficients $A_1,\dots,A_4$ for 90 integration paths $\Gamma_P^r$, where the distance from the crack tip $r$ ranges from $2\ \mathrm{mm}$ to $47\ \mathrm{mm}$. Additionally, we varied the number of integration points $P \in \{25, 50, 100, 200\}$ and the finite element mesh size $l_{\mathrm{e}} \in \{0.1, 0.2, 0.4, 0.8\}\ \mathrm{mm}$.

\begin{figure}
    \centering
    \includegraphics[width=\textwidth]{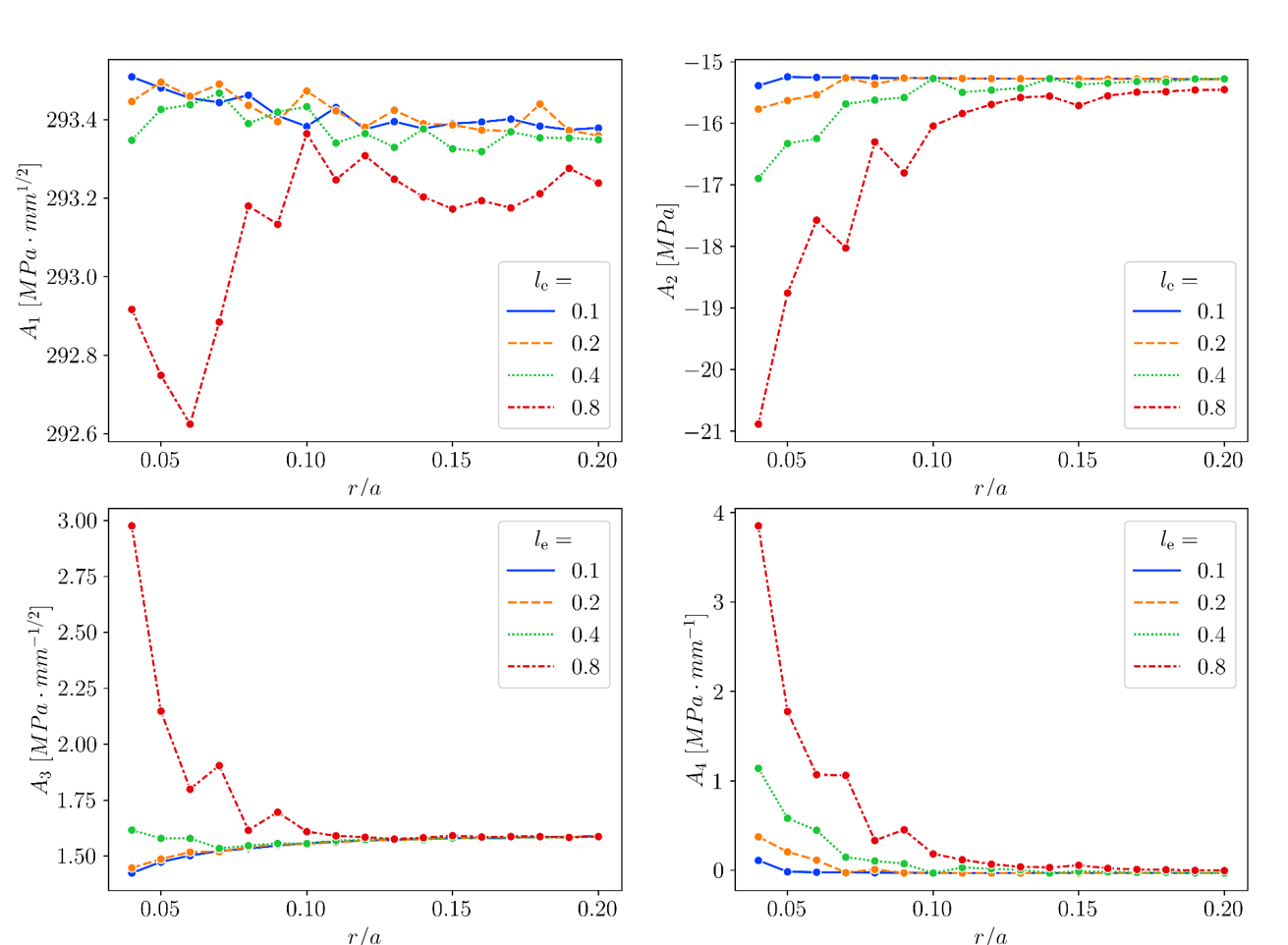}
    \caption{Convergence of $A_1,\dots,A_4$ for various crack tip distances of integration paths, different mesh sizes $l_{\mathrm{e}}$, and fixed number of integration points $P=100$.}
    \label{fig:conv_mesh_size}
\end{figure}

\begin{figure}
    \centering
    \includegraphics[width=\textwidth]{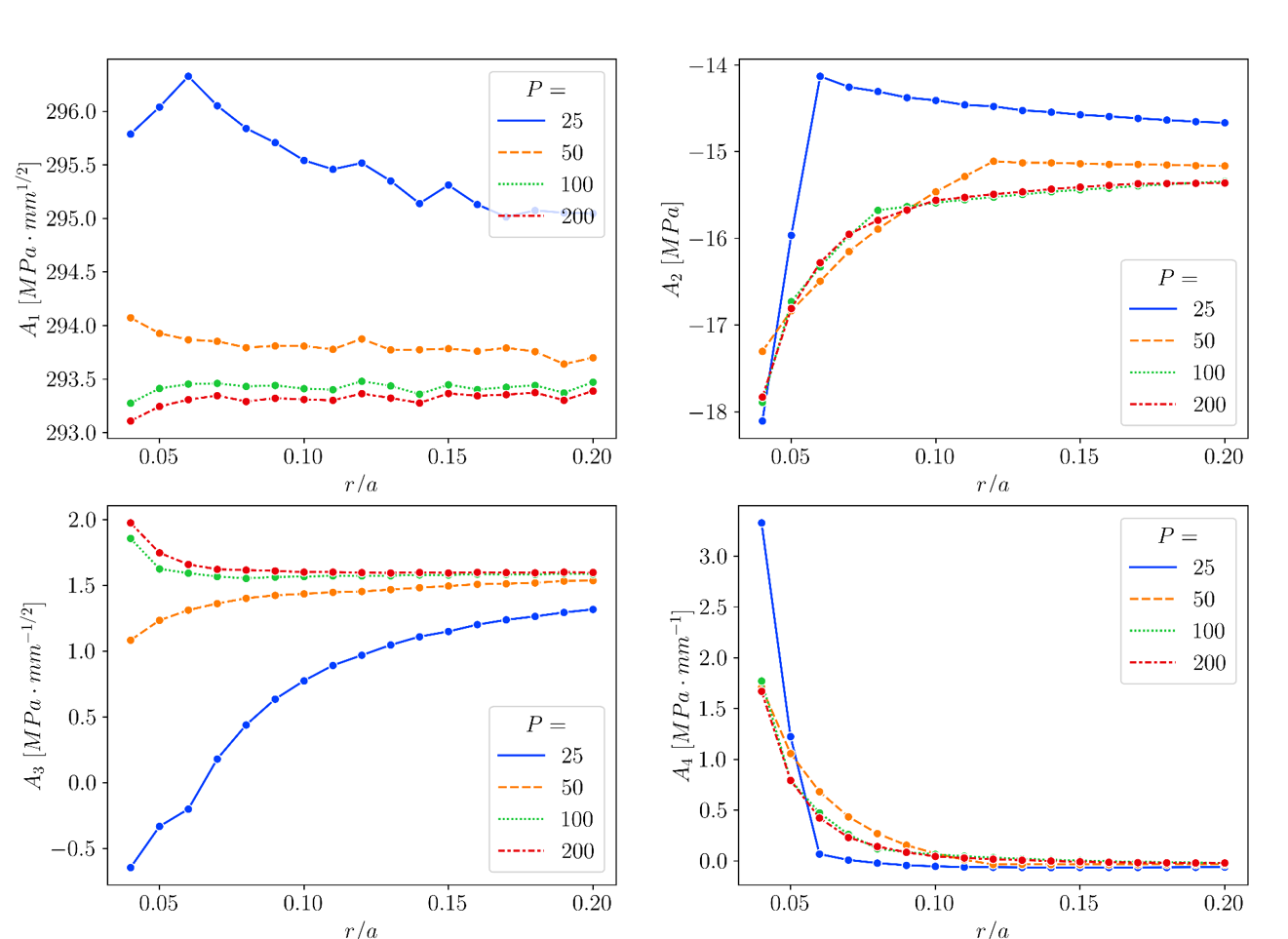}
    \caption{Convergence of $A_1,\dots,A_4$ for various crack tip distances of integration paths, different number of integration points $P$ and fixed$l_{\mathrm{e}}=1/2\ \mathrm{mm}$.}
    \label{fig:conv_num_nodes}
\end{figure}

Figure \ref{fig:conv_mesh_size} shows the results of $A_1,\dots,A_4$ for different mesh sizes $l_{\mathrm{e}}$ and in terms of the integral path depending on the relative crack tip distance. The number of integration points is fixed to $P=100$. In Figure \ref{fig:conv_num_nodes}, the same convergence is analyzed for fixed mesh size varying the number of integration points. In all cases, we observe exponential convergence in $r/a$ to a limit value. The speed of convergence increases with decreasing $l_{\mathrm{e}}$ and increasing $P$, i.e., the finer the domain and integral mesh the faster the convergence. Except for $P=25$, where the integral mesh is too coarse, all results have converged at a distance of $r/a=0.2$, which translates to $r=10\ \mathrm{mm}$.

Taking a more systematic approach, we denote the result for the $n$-th \textsc{Williams} coefficient from the integral path $\Gamma_P^r$, with radius $2\ \mathrm{mm} \le r \le 47\ \mathrm{mm}$, by $A_n^r$ and define the convergence distance as 

\begin{equation}
    r^\ast= {\arg \min}_s \left\{\max_{r>s}{\left|\frac{A_n^\infty-A_n^r}{A_n^\infty}\right|}<q\right\},
\end{equation}

where $q \in [0,1]$ is a convergence threshold and the limit $A_n^\infty$ is defined as the mean over $A_n^r$ for $r\ge 40\ \mathrm{mm}$ far away from the crack tip.

\begin{figure}
    \centering
    \includegraphics[width=\textwidth]{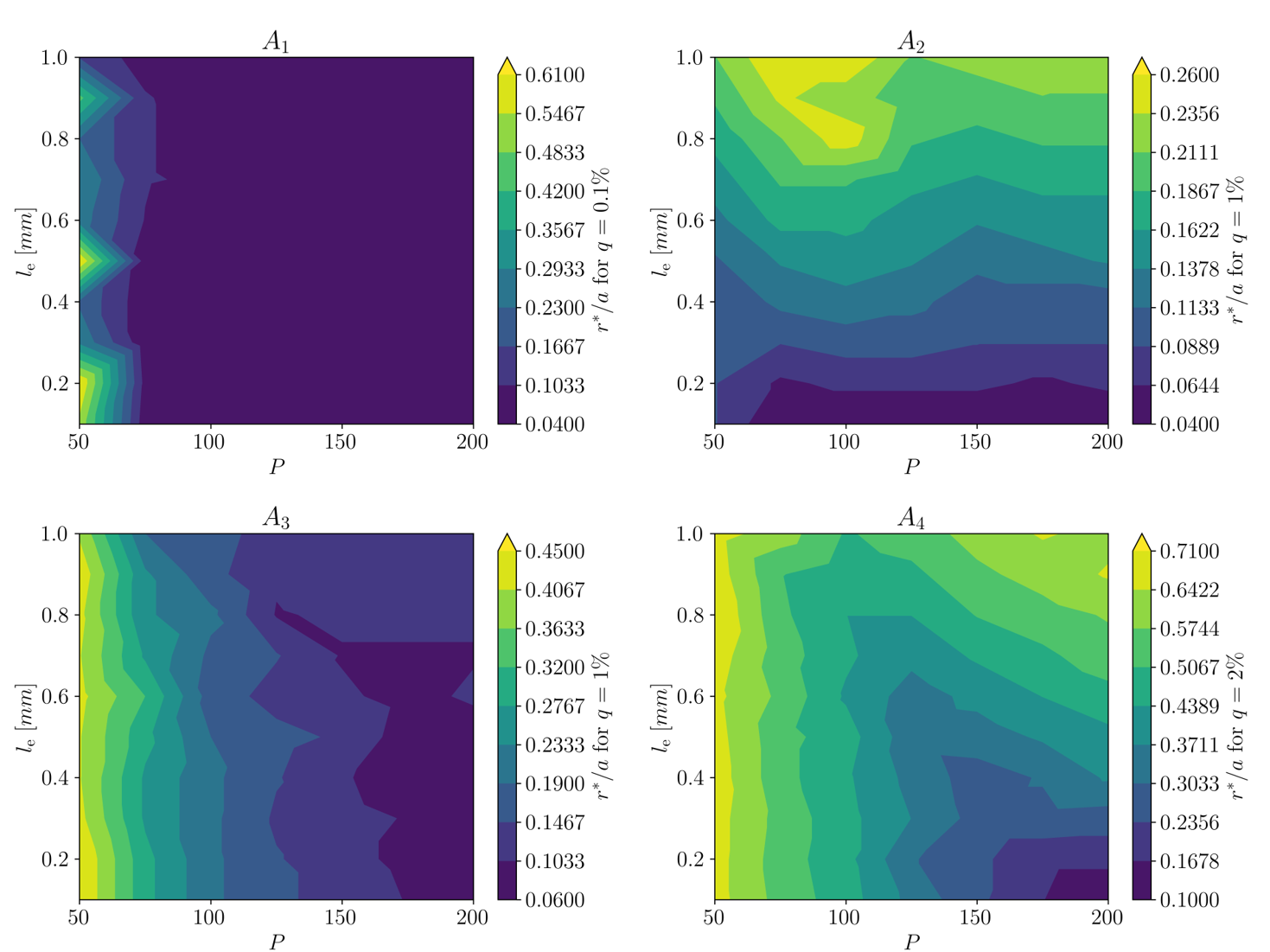}
    \caption{Convergence distance from the crack tip for $A_1,\dots,A_4$ depending on different $l_{\mathrm{e}}$ and $P$.}
    \label{fig:conv_normalized}
\end{figure}

In Figure \ref{fig:conv_normalized}, we see the (normalized) convergence distance $r^\ast/a$ for the first four mode-I terms of the \textsc{Williams} series expansion depending on $l_{\mathrm{e}}$ and $P$. In general, we observe that with higher order $n$ the convergence distance increases. For $A_1$ the convergence is almost instantaneous for $P>80$. For $A_2$ the convergence depends almost entirely on $l_{\mathrm{e}}$, whereas for $A_3$ it seems to depend solely on $P$. For $A_4$ we clearly observe faster convergence with smaller $l_{\mathrm{e}}$ and larger $P$. For visibility reasons, we used different thresholds $q$ for the four plots. 

We conclude that with a sufficiently fine mesh $l_{\mathrm{e}} \le 0.1\ \mathrm{mm}$ and sufficiently many integration points $P \ge 200$, the results for the terms $A_1,\dots,A_4$ for relative distances $r/a \ge 0.1$ have converged (up to a relative error of $q=2$ \%).

\subsection{Method comparison} \label{sec:results_method_comp}
In this section, we compare the integral method (INT) (see Sections \ref{sec:method_det_will_coeffs} and \ref{sec:method_int}) with the over-deterministic method (ODM) (see Section \ref{sec:method_odm}). We implemented the latter optimization method using the Python library \textit{scipy} \cite{2020SciPy-NMeth}. More specifically, we minimized the least squares error (see Equation \ref{eq:odm_minimization}) using the \textsc{Levenberg-Marquardt} iterative algorithm \cite{More77}. 

For the INT method, we use a mesh size of $l_{\mathrm{e}}=0.1$ for the finite element simulations and $P=200$ integration points for the integral approximation. We calculate 10 integrals between 8 and 10 mm distance from the crack tip ($0.16 \le  r/a \le 0.2$) and average the results. The convergence study above shows that the results have converged for this choice of parameters and crack tip distance (see Figure \ref{fig:conv_normalized}). For the ODM, a comparable fitting domain with inner radius $r_{\rm min}=8\ \mathrm{mm}$, outer radius $r_{\rm max}=10\ \mathrm{mm}$ and tick size $\delta= 0.03\ \mathrm{mm}$ is chosen.

\begin{figure}
    \centering
    \includegraphics[width=\textwidth]{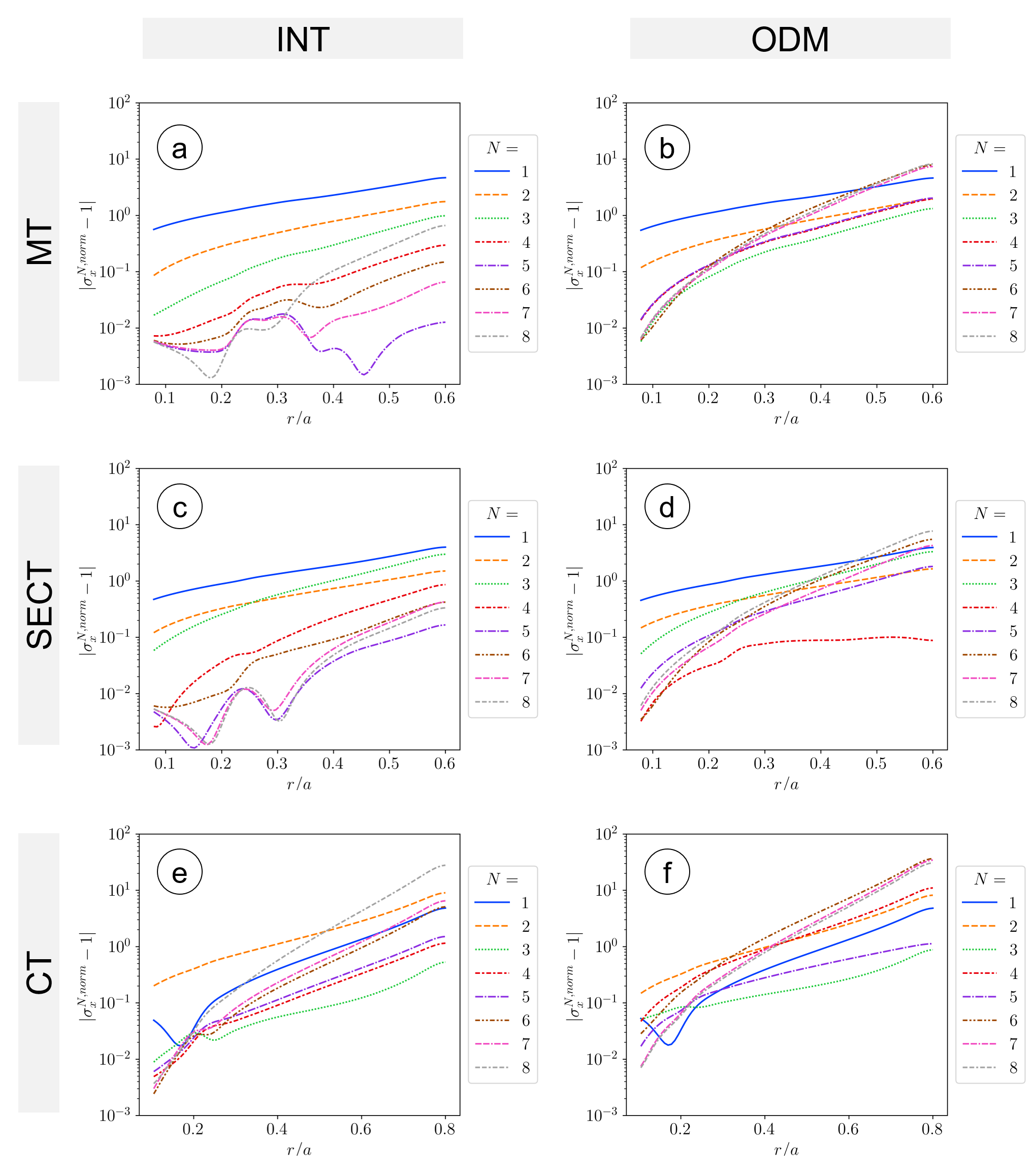}
    \caption{Relative approximation error of $\sigma_x$ for different \textsc{Williams} series expansion orders $N$ on the ligament $x_1>0$ ($\log$-scaling). The rows show the results for the MT- and SECT- and CT-specimen, respectively. The left and right column shows the results for the proposed integral method (INT) and the over-deterministic method (ODM), respectively.}
    \label{fig:approx_error}
\end{figure}

Figure \ref{fig:approx_error} shows the results of the relative approximation error $|\sigma_x^{N,\mathrm{norm}}-1|$, with $\sigma_x^{N,\mathrm{norm}} = \sigma_x^N / \sigma_x^{\mathrm{FE}}$, of the stress on the ligament in front of the crack tip. Here, $\sigma^{\mathrm{FE}}$ is the solution of the finite element simulation, and $\sigma^{N}$ is the truncated \textsc{Williams} stress (cf. Equation \ref{eq:williams_stress_expansion}). We compare INT and ODM for all three specimens at a crack length of $\alpha=0.5$. The number of coefficients taken into account varies from $N=1,\dots,8$. 

Overall, both methods almost coincide up to order $N=3$, but the INT method consistently shows a much lower relative approximation error than ODM for $N>3$. In that case, especially for the MT- and SECT-specimen the INT method is orders of magnitude more precise than ODM. For both methods and all crack tip distances, the approximation error first decreases when more \ac{HORTs} are taken into account. However, at a certain number of terms the approximation gets worse. Focusing on the MT-specimen, we observe the best approximation for INT at order $N=5$ at a relative error less than $10^{-2}$ (consistently for all distances $r/a$), whereas for ODM the order $N=3$ shows the lowest error rising from order $10^{-2}$ to order $1$ with increasing crack tip distance. For SECT and CT, the INT method best approximates the stress field for $N=5$ and $N=3$, respectively.

\subsection{Parameter study} \label{sec:results_param_study}
In order to demonstrate the integral method, we now study the influence of the parameters $F,a,W,$ and $H$ on \ac{HORTs} for the three specimen types introduced above (see Section \ref{sec:method_simulations}). The goal is to generate a rich database to analyze the influence on the crack tip field characterized by the \textsc{Williams} coefficients.

For each specimen type, the parameter study was carried out in two steps. First, we implemented the model in Ansys\textsuperscript{\textregistered} and used pyAnsys \cite{pyAnsys} to setup, solve, and export nodal displacements and strains for each parameter configuration. Second, the \textsc{Williams} coefficients were calculated using INT implemented as described in Sections \ref{sec:method_det_will_coeffs} \& \ref{sec:method_int} above as part of \textit{crackpy} \cite{crackpy-github}.

For this, we use a mesh size of $l_{\mathrm{e}}=0.1\ \mathrm{mm}$ for the finite element simulations and $P=200$ integration points. We calculate 10 integrals between 8 and 10 mm distance from the crack tip and average the results. The convergence study above shows that the results have converged for this choice of parameters and crack tip distance (see Figure \ref{fig:conv_normalized}). 

To plot several coefficients together, it is useful to normalize the \textsc{Williams} coefficients $A_n$ to dimensionless coefficients $a_n$ as follows \cite{Karihaloo.2003}:
\begin{equation} \label{eq:williams_normalized}
    a_n=\frac{A_n}{\sigma_{\rm nom}W^{1-n/2}},
\end{equation}
where $\sigma_{\rm nom}=Ft^{-1}W^{-1}$ denotes the nominal stress induced by the external force on the upper boundary.

\subsubsection{MT-specimen} \label{sec:results_param_study_MT}
For the MT-specimen, we varied 
\begin{itemize}
    \item $H\in \{100, 200, 300, 400\}\ \mathrm{mm}$
    \item $W\in \{120, 160, 200\}\ \mathrm{mm}$
    \item $\alpha \in \{0.2, 0.25, 0.3, 0.4, 0.45, 0.5, 0.55, 0.6, 0.65, 0.7, 0.75, 0.8\}$
    \item $F\in \{2.5, 5, 7.5, 10\}\ \mathrm{kN}$
\end{itemize}

As theoretically expected, the force enters linearly in all the \textsc{Williams} coefficients and does not change the normalized coefficients $a_n$. 

\begin{figure}
     \centering
     \begin{subfigure}[b]{0.49\textwidth}
         \centering
         \includegraphics[width=\textwidth]{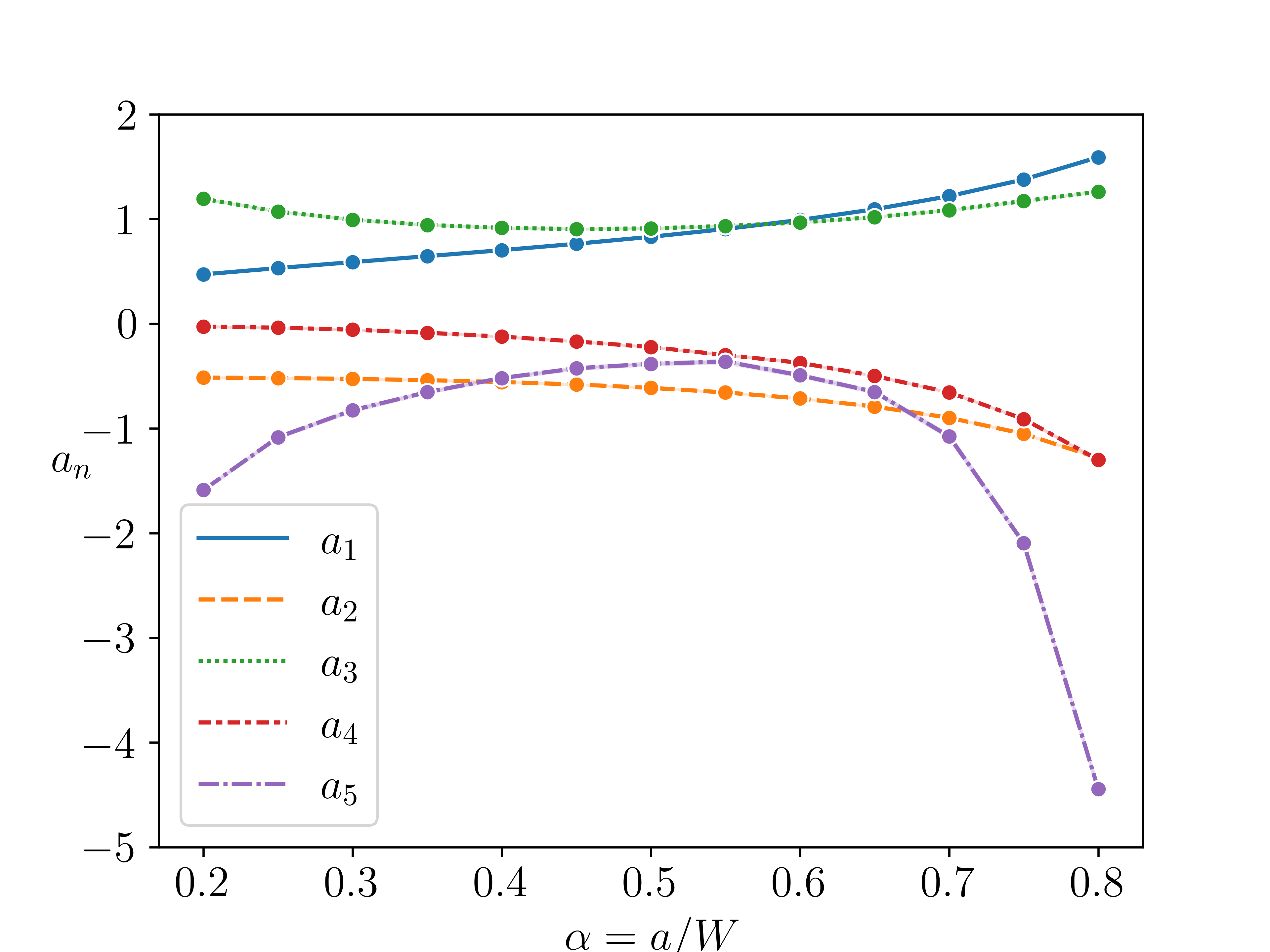}
         \caption{$H/W=1$}
         \label{fig:MT_param_study_HW1}
     \end{subfigure}
     \hfill
     \begin{subfigure}[b]{0.49\textwidth}
         \centering
         \includegraphics[width=\textwidth]{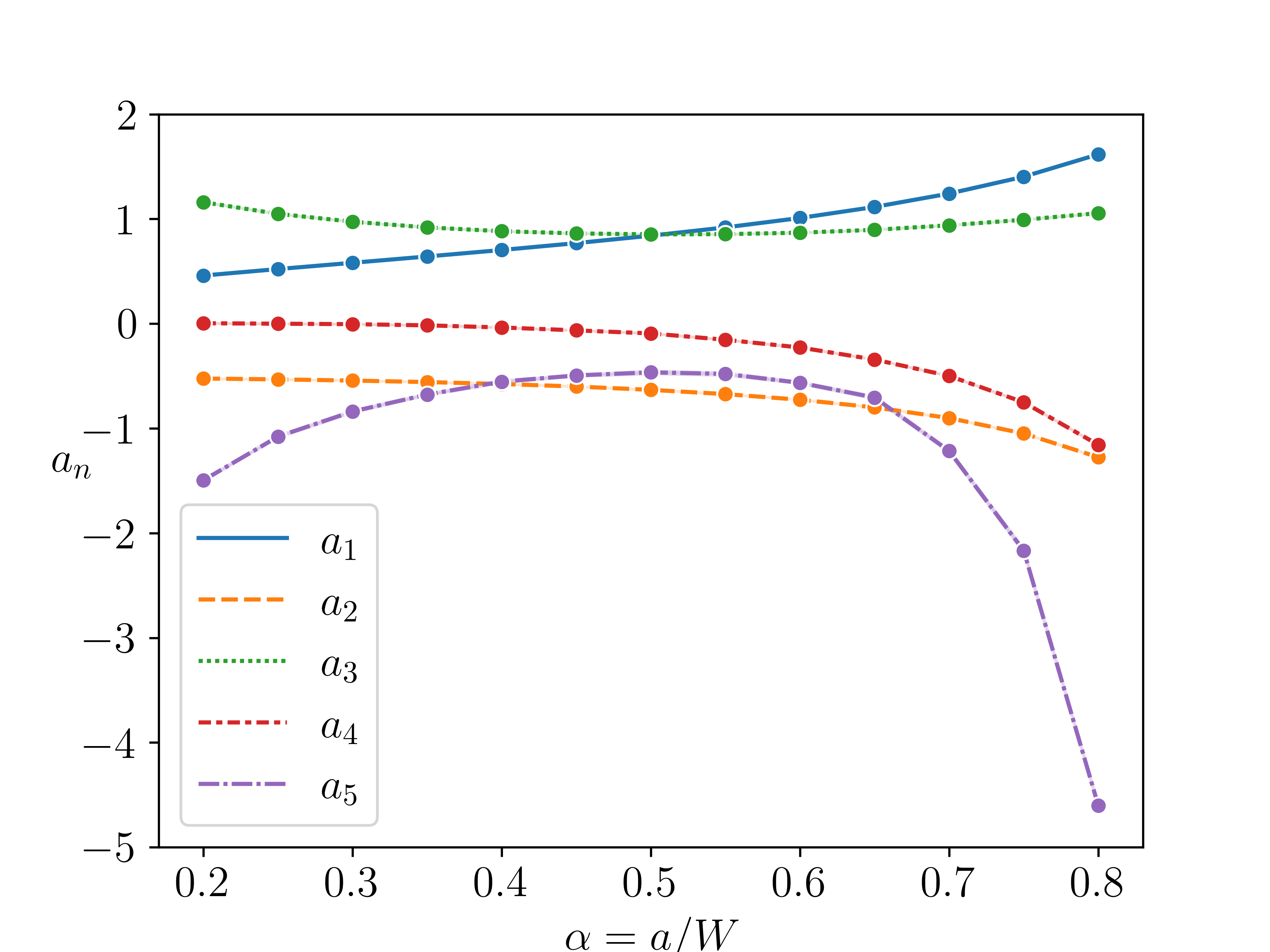}
         \caption{$H/W=2$}
         \label{fig:MT_param_study_HW2}
     \end{subfigure}
    \caption{Normalized \textsc{Williams} coefficients $a_1,\dots,a_5$ for the MT-specimen parameter study over the whole crack length $\alpha$.}
    \label{fig:MT_param_study}
\end{figure}

Figure \ref{fig:MT_param_study} shows the first five normalized \textsc{Williams} coefficients depending on the crack length for two different MT-specimen geometries. We observe a positive $T$-stress throughout the crack growth with only marginal influence of the geometry descriptor $H/W$ on the coefficients.

\subsubsection{CT-specimen} \label{sec:results_param_study_CT}
For this specimen, we also fixed the force $F=10\ \mathrm{kN}$ and varied
\begin{itemize}
    \item $W\in \{75, 80, 85, 90, 95, 100\}\ \mathrm{mm}$
    \item $\alpha \in \{0.2, 0.25, 0.3, 0.4, 0.45, 0.5, 0.55, 0.6, 0.65, 0.7, 0.75, 0.8\}$
\end{itemize}

\begin{figure}
     \centering
     \begin{subfigure}[b]{0.49\textwidth}
         \centering
         \includegraphics[width=\textwidth]{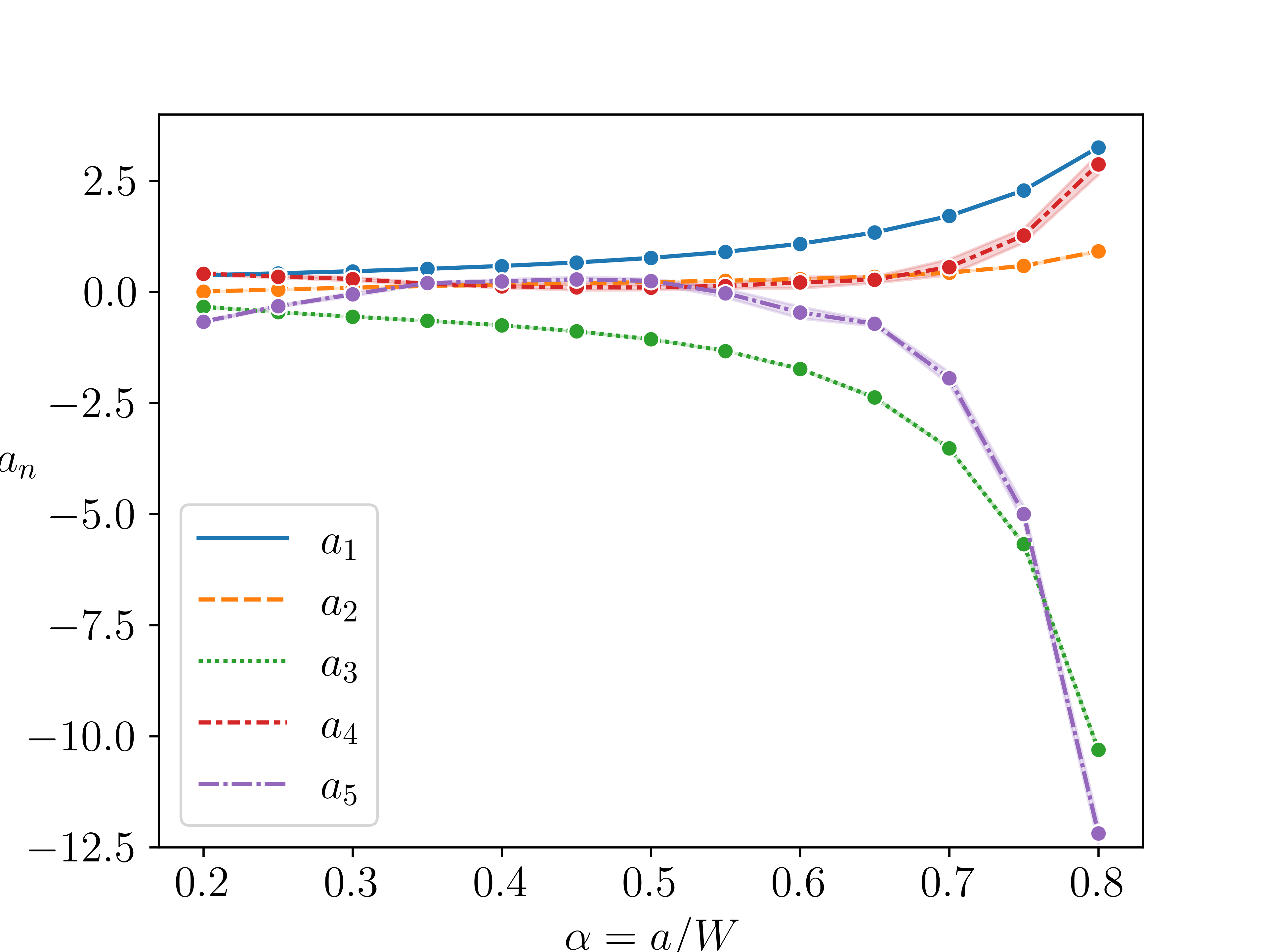}
         \caption{Normalized coefficients for $W=100$}
         \label{fig:CT_param_study_W100}
     \end{subfigure}
     \hfill
     \begin{subfigure}[b]{0.49\textwidth}
         \centering
         \includegraphics[width=\textwidth]{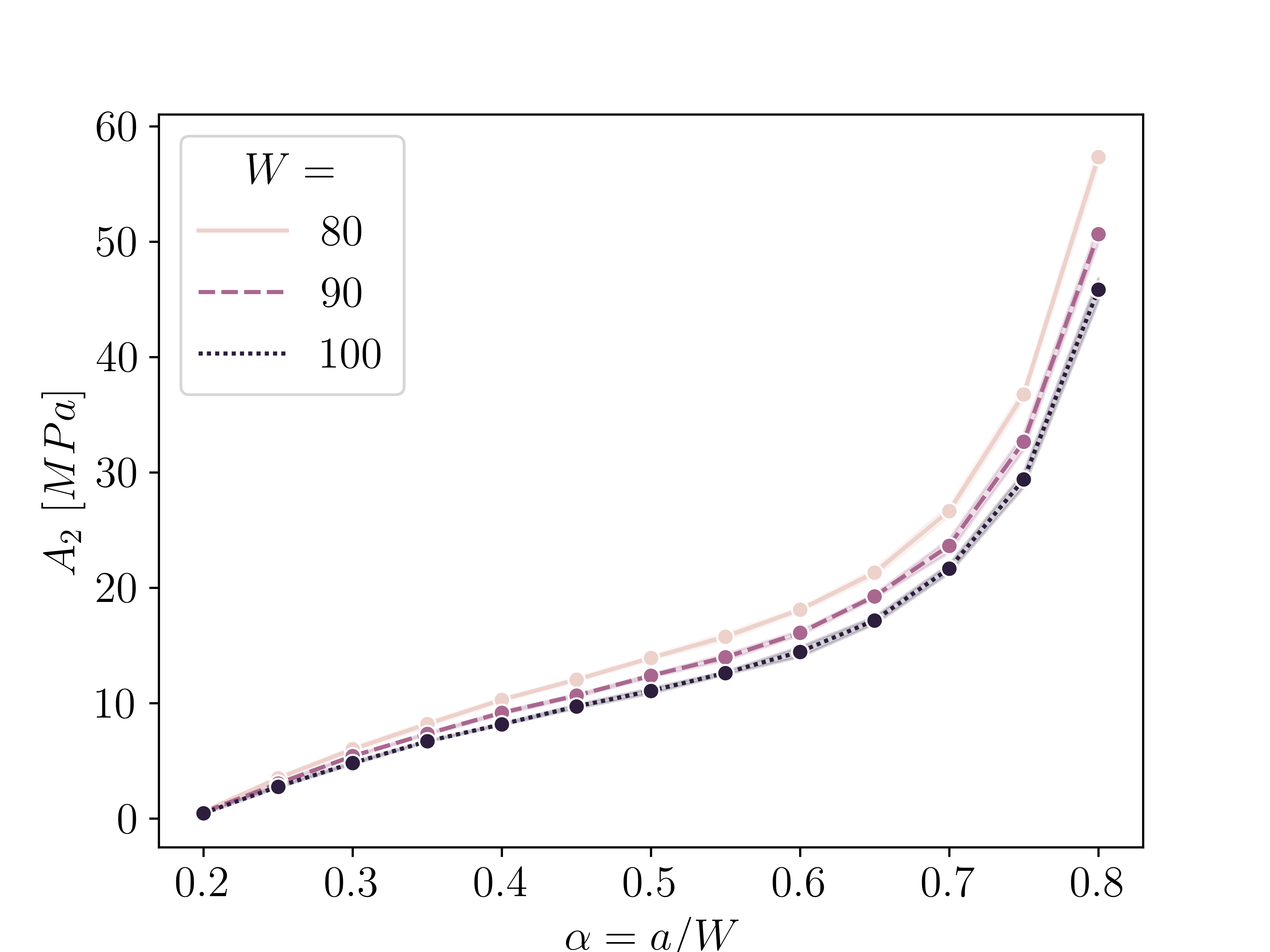}
         \caption{$A_2$ for different widths}
         \label{fig:CT_param_study_T_stress}
     \end{subfigure}
    \caption{Normalized \textsc{Williams} coefficients $a_1,\dots,a_5$ for the CT-specimen parameter study over the whole crack length $\alpha$ for $W=100$ (a) and \textsc{Williams} coefficient $A_2=T/4$ for different widths (b).}
    \label{fig:CT_param_study}
\end{figure}

Note that the $H/W$-ratio is fixed to 1.2 for the CT-specimen.
The results are shown in Figure \ref{fig:CT_param_study}. In contrast to the previous parameter studies, we observe a positive $T$-stress for all crack lengths rising nonlinearly up to $200\ \mathrm{MPa}$. As $T$-stress is related to the directional stability of the crack path, this finding suggests the possibility of crack path deflections in actual experiments with such CT samples. The third coefficient controls the stability of the crack propagation \cite{Cotterell1966}. Therefore this coefficient can be related to unstable crack propagation which is of particular importance. For the parameter study of the MT-specimen, the third coefficient is positive and fairly stable throughout the crack growth. However, for the CT-specimen it is always negative and substantially decreases even more when $\alpha > 0.7$.

\subsubsection{SECT-specimen} \label{sec:results_param_study_SECT}
In this parameter study, we now fixed the force $F=10\ \mathrm{kN}$ and varied 
\begin{itemize}
    \item $H \in \{50, 100, 300, 400\}\ \mathrm{mm}$
    \item $W \in \{100, 150\}\ \mathrm{mm}$
    \item $\alpha\in \{0.2, 0.25, 0.3, 0.4, 0.45, 0.5, 0.55, 0.6, 0.65, 0.7, 0.75, 0.8\}$
\end{itemize}

\begin{figure}
     \centering
     \begin{subfigure}[b]{0.49\textwidth}
         \centering
         \includegraphics[width=\textwidth]{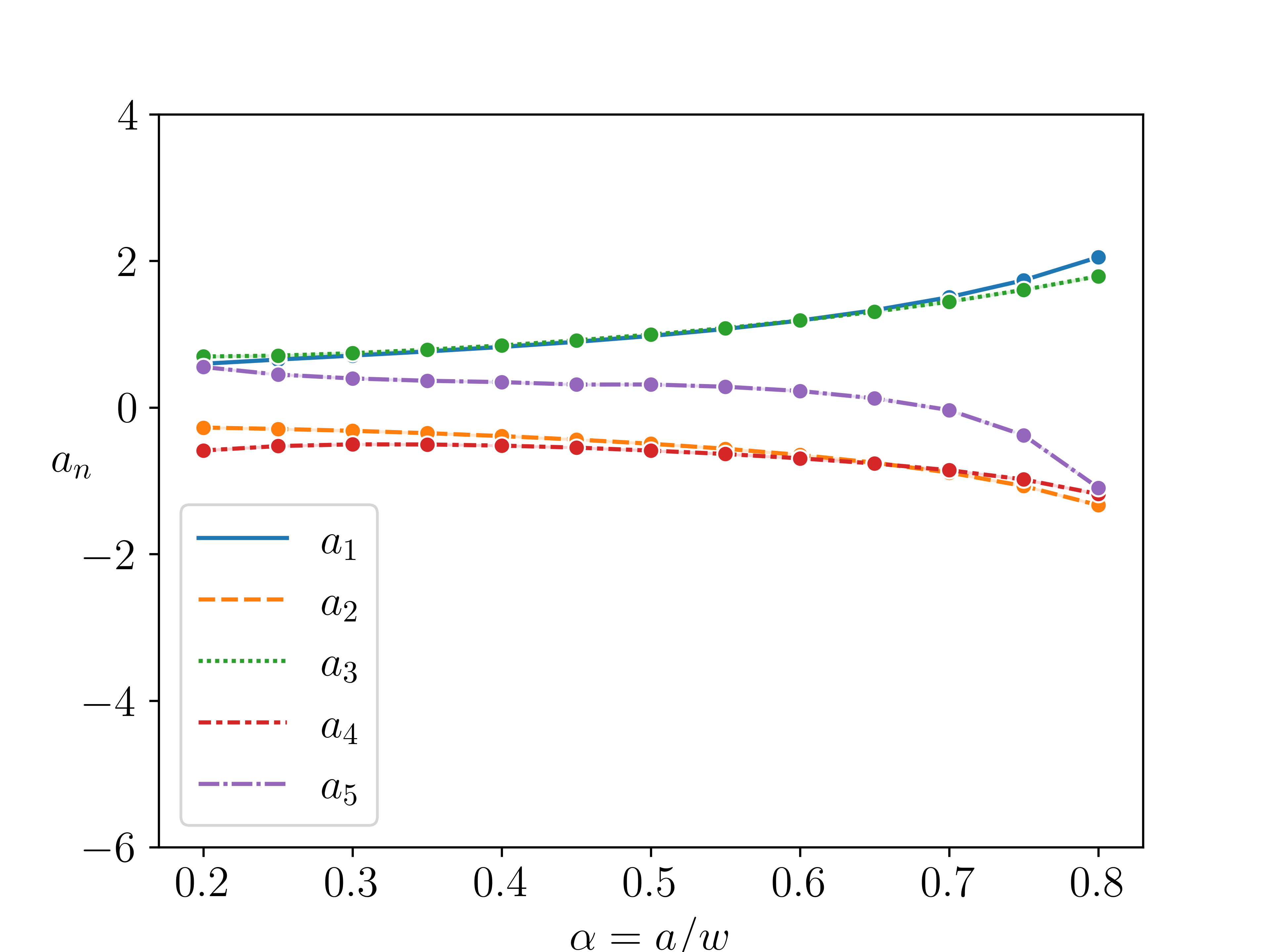}
         \caption{$H/W=1$}
         \label{fig:SECT_param_study_HW1}
     \end{subfigure}
     \hfill
     \begin{subfigure}[b]{0.49\textwidth}
         \centering
         \includegraphics[width=\textwidth]{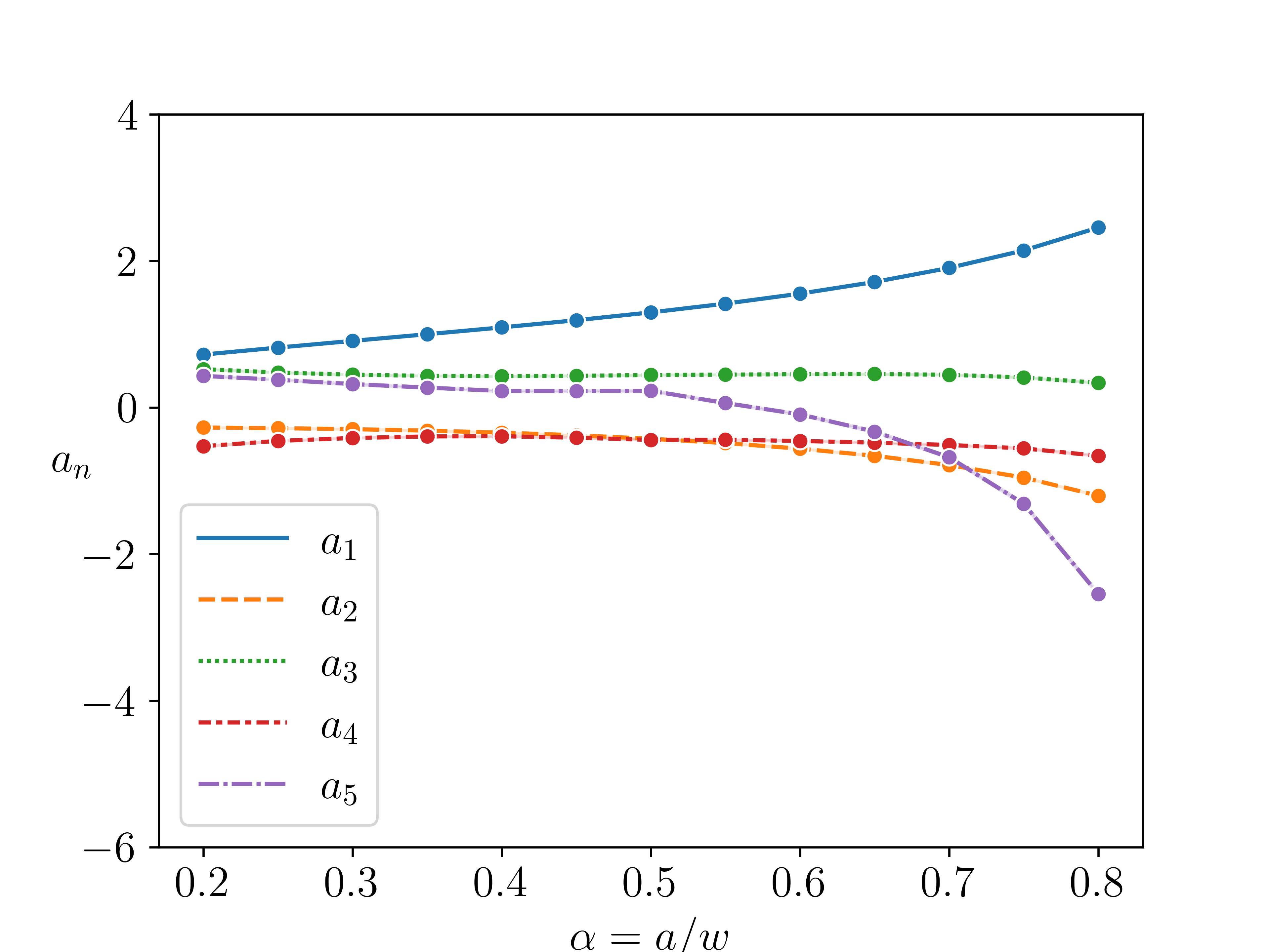}
         \caption{$H/W=2$}
         \label{fig:SECT_param_study_HW2}
     \end{subfigure}
     \hfill
          \begin{subfigure}[b]{0.49\textwidth}
         \centering
         \includegraphics[width=\textwidth]{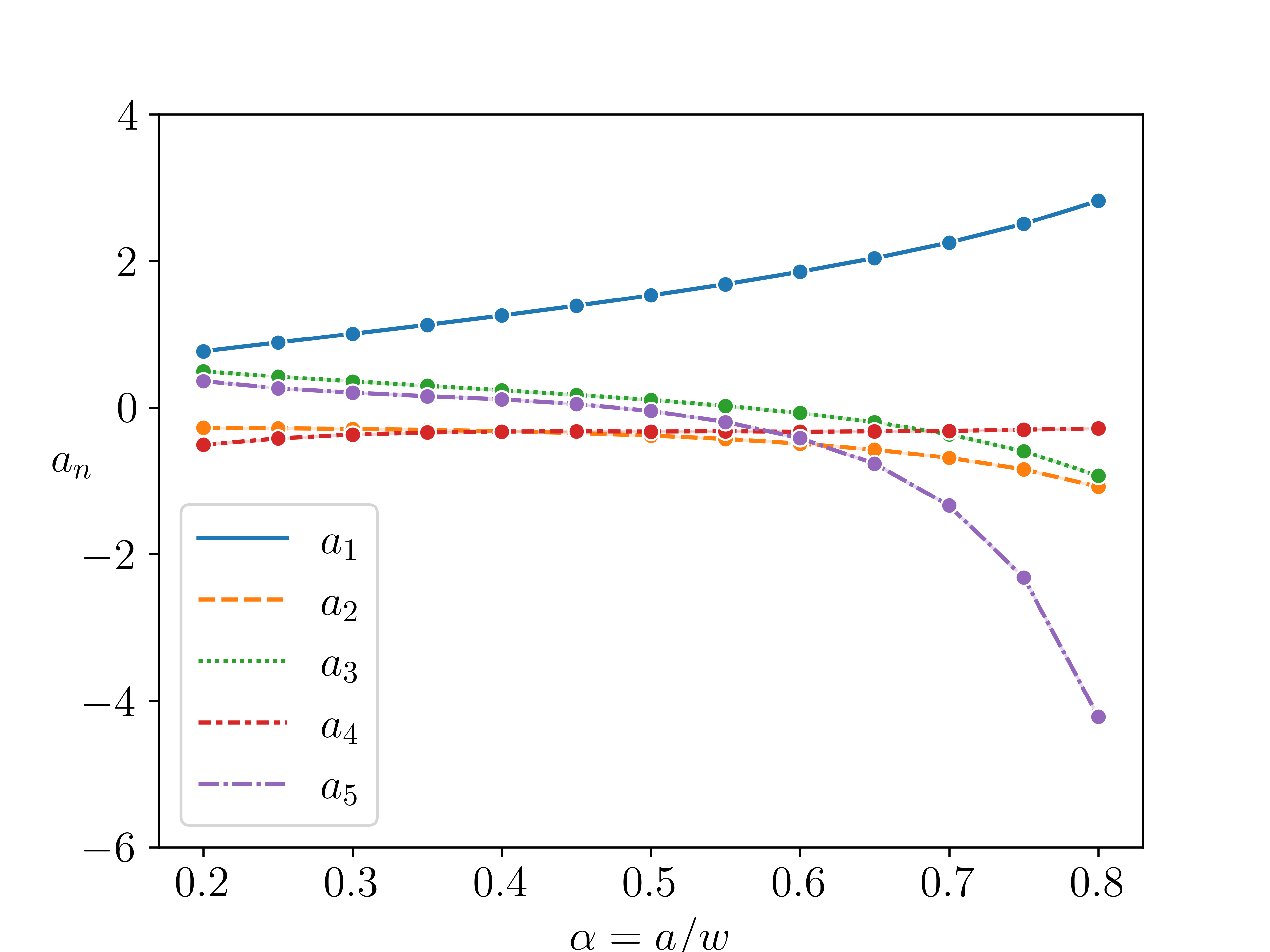}
         \caption{$H/W=3$}
         \label{fig:SECT_param_study_HW3}
     \end{subfigure}
     \hfill
     \begin{subfigure}[b]{0.49\textwidth}
         \centering
         \includegraphics[width=\textwidth]{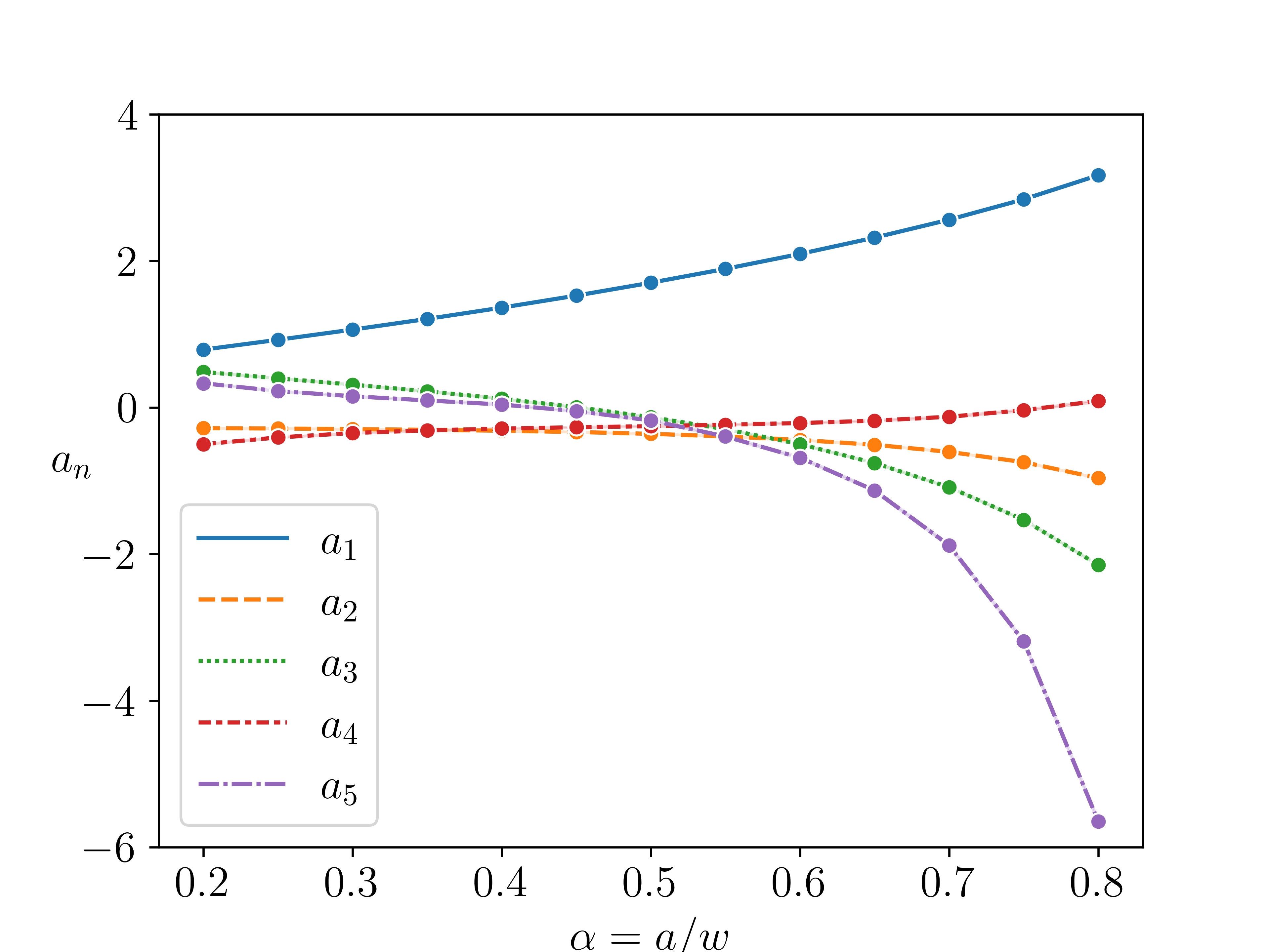}
         \caption{$H/W=4$}
         \label{fig:SECT_param_study_HW4}
     \end{subfigure}
     
    \caption{Normalized \textsc{Williams} coefficients $a_1,\dots,a_5$ for the SECT-specimen parameter study over the whole crack length $\alpha$ and for several $H/W$ ratios.}
    \label{fig:SECT_param_study}
\end{figure}

In contrast to the MT-study above, the calculated \textsc{Williams} coefficients and especially the \ac{HORTs} clearly depend on the geometry of the domain in terms of the $H/W$-ratio (Figure \ref{fig:SECT_param_study}). In particular, the results of the third term, which is related to the stability of crack propagation, are interesting. The third term develops completely different for the two geometries in Figure \ref{fig:SECT_param_study}. Whereas for $H/W=1$ the third term is always positive and increases during crack growth, for $H/W=3$ and $H/W=4$ it decreases and gets even negative.

\section{Conclusion} \label{sec:conclusion}

We introduced a method to determine coefficients of the famous \textsc{Williams} series known to describe the singular crack tip field in plane elasticity. The method is based on \textsc{Bueckner}'s conjugate work integral and was first described by \textsc{Chen} \cite{Chen1985}. 
We studied the convergence of path integral results in terms of the distance from the crack tip and compared the integral method to the over-deterministic method (ODM), which is based on fitting the Williams series expansion (up to a certain term $N$) to measured or simulated displacement data.
We carried out an exhaustive parameter study for the different specimen types in order to generate a rich data basis for further analysis.

Considering the results in Section \ref{sec:results}, we draw the following conclusions: 

\begin{itemize}

    \item Calculating higher order terms too close to the crack tip leads to unstable results due to numerical instabilities. With higher order, the distance needed for convergence increases and further depends on the integral and domain mesh size. To correctly calculate the first four terms, we found that at a domain mesh size smaller than $0.1\ \mathrm{mm}$ a relative distance of $r/a\ge0.1$ and more than $200$ integration points are needed. 
    
    \item With \ac{ODM}, modifying the total number of terms $N$, alters the results for all coefficients. This property makes it difficult to decide how many terms are actually needed. In contrast, coefficients determined by the integral method are independent of the total number of coefficients taken into account since every coefficient is calculated separately.
    
    \item Coefficients calculated with the integral method describe the crack tip field more precisely. Throughout all specimen types, this result manifest itself in lower stress field approximation errors compared to the \ac{ODM}.
    
\end{itemize}

\section{Outlook}
From a data scientific standpoint, the precise calculation of \textsc{Williams} coefficients is a dimension reduction. It gives an efficient way to compress the high-dimensional nodal displacement data ($\approx10^5$ D) of the crack tip field in the much lower dimensional space of coefficients ($\approx10$ D).
This approach opens the possibility to search for patterns and correlations which influence crack path and crack propagation stability. It hopefully paves the way towards a modern data-driven and theory-guided investigation of crack propagation.
The present work focuses on linear-elastic simulations of mode-I crack configurations. For future work, it would be interesting to investigate the ability of this method to capture plastic effects (e.g. by using elastic-plastic simulations). Further, the method should be tested and analyzed for real experimental data.

\section{Acknowledgements}
The authors gratefully acknowledge the financial support of the DLR Aeronautics Directorate.
\printacronyms

\bibliographystyle{unsrt}
\bibliography{sample}

\end{document}